\documentclass[twocolumn]{aastex631}

\usepackage{ifluatex}
\ifluatex
\usepackage{pdftexcmds}
\usepackage{hyperref}
\makeatletter
\let\pdfstrcmp\pdf@strcmp
\let\pdffilemoddate\pdf@filemoddate
\makeatother
\fi

\usepackage{comment}
\graphicspath{{./}{figures/}}

\begin{document}

\title{Magnetic diagnostics of prominence eruptions through the Hanle effect of the \ion{He}{1} 1083\,nm line}

\correspondingauthor{Momchil E. Molnar}
\email{mmolnar@ucar.edu}
\author[0000-0003-0583-051X]{Momchil E. Molnar}
\affiliation{High Altitude Observatory, 
National Center for Atmospheric Research, Boulder, CO}

\author[0000-0001-6990-513X]{Roberto Casini}
\affiliation{High Altitude Observatory, 
National Center for Atmospheric Research, Boulder, CO}



\begin{abstract}

The magnetic field vector of the solar corona is not regularly and comprehensively being measured, because of
the complexity and degeneracy inherently present in the types of observations currently available.
To address some of the current limitations of coronal polarimetry, we present computations that demonstrate the possibility of 
magnetometry using the unsaturated Hanle effect of the \ion{He}{1} 1083\,nm line. 
The main purpose of this investigation is to show how the geometric properties of the linear
polarization of this line can be used to routinely diagnose the orientation of the field in erupting prominences, thus providing an important 
constraint on the $B_z$ determination at 1\,AU. For this work, we adopted a simplified magnetic  model 
of a flux rope, consisting of a toroidal helical structure embedded in a hydrostatically stratified corona.
Our results demonstrate the possibility
to discern different orientations of the  magnetic field 
vector in such structures under rather general and practicable viewing conditions. In particular, observations from the Sun-Earth
Lagrange points are found to provide excellent locations for the deployment of synoptic 
instruments aiming at the estimation of the magnetic field of Earth-directed Coronal Mass Ejections. We complete our demonstration by showing how a small (${\sim}5$\,cm) 
space-borne coronagraph 
can achieve sufficient signal-to-noise ratios to make the 
coronal magnetometry goal outlined above feasible.

\end{abstract}



\section{Introduction} 
\label{sec:intro}

In order to understand the physical processes driving 
coronal dynamics and heating we need to further our understanding 
of the coronal magnetic field 
\citep[see, e.g., the review by][and 
references therein]{2019ARA&A..57..157C}. 
Furthermore, to improve the space-weather prediction capabilities with regards to
the geo-effectiveness of Earth-directed solar eruptions,
we need to observe their magnetic structure
\citep{2019RSPTA.37780096V}, which is currently not possible.
Inferring the coronal magnetic field is challenging because of the very low contrast
(${\sim}1$ to 10\,ppm) of coronal structures to disk brightness.

Recent efforts to measure the coronal magnetic field vector rely on different
physical mechanisms \cite[see, e.g.,][]{2017SSRv..210..145C}, each of them suffering from various limitations.
The longitudinal Zeeman 
effect in forbidden emission lines of the corona has been used to estimate coronal magnetic fields above active regions
\citep{2000ApJ...541L..83L,2004ApJ...613L.177L}.
These pioneering measurements have shown that such fields range between a few to a few tens of gauss. 
These coronal fields
produce small circular polarization amplitudes ($\sim 10^{-4}$ to $10^{-3}$ of the line intensity; 
\citealt{2000ApJ...541L..83L,2004ApJ...613L.177L}), making these diagnostics particularly challenging from an instrumental point of view~\citep{Schad_2024}. In addition, the linear polarization of these lines is completely dominated by resonance scattering, and practically insensitive to the magnetic field strength  \cite[e.g.,][]{2017SSRv..210..145C}. Thus, no vector diagnostics of the coronal magnetic field is generally possible with these lines, without further constraints \citep{2014ApJ...792...23P,2021ApJ...912...18J,2022SoPh..297...63P}.

This situation can be ameliorated if additional magnetic effects in the plasma can be observed.
As an example, the Upgraded Coronal Multi-Channel Polarimeter (UCoMP; \citealt{2019shin.confE.131T}) 
has successfully been used to derive magnetic maps of the 
plane-of-sky (POS) projected component of the field based on the study of
Alfv\'{e}nic wave propagation in the corona. This method was able to confirm typical strengths of the
coronal magnetic field of a few gauss \citep{2020ScChE..63.2357Y,2020Sci...369..694Y}.
On the other hand, radio measurements have indicated that the magnetic field over flaring regions could be on the order of hundreds of gauss \citep{2023ApJ...943..160F,2022Natur.606..674F}.
This wide range of magnetic field strengths
inferred using different techniques is still a subject of ongoing debate.

Another proposed method to infer the magnetic field in the solar corona is through the
polarimetric signatures of permitted transitions that are sensitive to the Hanle effect, 
over ranges of field strength such as those found in the corona \citep{2016FrASS...3...20R,2022SoPh..297...96K,2024arXiv240605539K}. 

The Hanle effect  \citep{1924ZPhy...30...93H} is a quantum mechanical 
phenomenon where atomic-level coherence induced by anisotropic and/or polarized
incoming radiation is partially relaxed by the presence of a magnetic field, resulting in an observable
modification of the 
linear polarization signals (through a change of polarization degree and a rotation of the 
polarization plane) in the scattered radiation, with respect to
the same signals in the absence of a magnetic field. 
The Hanle effect is sensitive to the full vector of the magnetic field, over a range of strengths that extends around a critical field value determined by the 
inverse lifetime of the atomic level,
and how this compares with the Larmor ``precession'' frequency associated with the ambient field. The 
usable range of field strengths varies approximately between 1/10 to 10 times this critical field 
value \cite[e.g.,][]{2004ASSL..307.....L}. In this range, called the ``unsaturated'' Hanle regime, the linear polarization shows the largest 
rate of change with the magnetic field. When the magnetic strength far exceeds the regime of sensitivity to the Hanle effect, the phenomenon reaches a state of ``saturation'',
where atomic-level coherence is fully relaxed, and the polarization degree and orientation become independent of the field strength. This saturation condition is the reason why the linear polarization of forbidden emission lines in the corona is insensitive to the magnetic strength of typical coronal fields, 
as the very low probability of spontaneous decay of the excited atomic levels corresponds to critical fields typically lower than a microgauss.

\citet{1999A&A...345..999R} and \citet{2002A&A...386..721R} 
used the signatures of the unsaturated Hanle effect in
the corona from SUMER/SoHO observations \citep{1995SoPh..162..189W}. 
Magnetic field strengths on the order of a few gauss above 
the solar poles, and an outflow speed of a few tens of 
km/s of the young solar wind, were found based on this
approach \citep{2002A&A...396.1019R}. Recent efforts 
by \cite{2019ApJ...883...55Z,2021ApJ...920..140S,2022SoPh..297...96K,2024arXiv240605539K},
have identified FUV and EUV spectral windows containing multiple permitted transitions
with critical field values ranging between 10$^{-3}$\,G and 10$^2$\,G, thus providing a comprehensive set
of potential diagnostics to cover the different magnetic field regimes observed (and expected) in the corona.

In this paper we illustrate how off-limb observations of the
unsaturated Hanle effect in the \ion{He}{1} 1083.0 nm 
multiplet can readily provide an estimate of the coronal magnetic field strength and its orientation.
This idea has been proposed previously by \citet{2016FrASS...3...20R} and \citet{2016FrASS...3...13D},
as the Hanle magnetic sensitivity range of the \ion{He}{1} 1083 line is 
approximately 0.1-10\,gauss, which is optimally suited for the expected range of coronal field strengths.
However, a detailed study of the robustness of this diagnostic, based on realistic
magnetic topologies of the solar corona is lacking in the literature. 

\ion{He}{1} 1083 coronal signals have previously been detected
during solar eclipses \citep{1996ApJ...456L..67K,Dima2018},
and using large aperture coronagraphs \citep{2007ApJ...667L.203K,2010ApJ...722.1411M}. 
Detailed calculations by 
\citet{2020ApJ...898...72D} of the expected brightness under coronal conditions also support the case that the \ion{He}{1} 1083\,nm line should be comparable in 
brightness to the well-known diagnostic pair of \ion{Fe}{13} 1075/1080\,nm forbidden coronal 
lines. 

Recent reanalyses of past data have challenged the coronal origin of the \ion{He}{1} signals observed in the diffuse corona (Dima 2024, private communication). For this reason, this work mainly focuses on the polarization
signatures of \ion{He}{1} in prominence eruptions, since prominences carry 
a significant amount of chromospheric plasma  
strongly radiating in \ion{He}{1} 1083\,nm, as seen for example in the UCoMP and  
CHIP instrument archives.\footnote{\href{https://www2.hao.ucar.edu/mlso}{MLSO archive, including the UCoMP and CHIP data archives} ~\citep{2019shin.confE.131T, Kopp1997}.} Furthermore, recent observations with the METIS coronagraph on 
board Solar Orbiter \cite[SO;][]{2020A&A...642A..10A} suggest that polarized emission in the
\ion{He}{1} D$_3$ 587.6\,nm multiplet was detected in a prominence eruption up to 9\,$R_\odot$ \citep{2023ApJ...957L..10H}. Because \ion{He}{1} 1083 and D$_3$ share the first excited ${}^3$P$_{0,1,2}$ 
atomic term of orthohelium, these SO observations further corroborate the interest of
\ion{He}{1} 1083 for the magnetic diagnostics of
coronal mass ejections (CMEs), up to very large distances from the Sun. 

Based on such compelling
theoretical and observational evidence, we believe that the present study will help fill a critical
knowledge gap in the applicability of the unsaturated Hanle effect for coronal magnetism diagnostics. In Section~\ref{sec:Numerical_setup}
we describe the radiative transfer code, the \ion{He}{1} atomic model, and the coronal 
flux-rope model, which we adopted for this study. Using this numerical setup, we
compute the polarization signatures from a flux rope observed above the solar limb for different
viewing geometries (Section~\ref{sec:Results_observables}). 
In particular, we model polarimetric 
observations from the Sun-Earth 
Lagrange points L4 and L5 of a flux rope located along the Sun-Earth line, in order to demonstrate
the unique capability for predicting the magnetic field structure of Earth-directed, interplanetary CMEs (ICMEs) from those viewpoints. Finally, in Section~\ref{sec:Discussion} we provide estimates of the expected 
signal-to-noise ratios of coronal
\ion{He}{1} 1083 observations assuming a small aperture (D\,$\sim 5$\,cm) space-borne coronagraph. 

\section{Numerical setup}
\label{sec:Numerical_setup}

\begin{figure*}[ht!]
\begin{minipage}[c]{0.325\textwidth}
	\includegraphics[width=\textwidth]{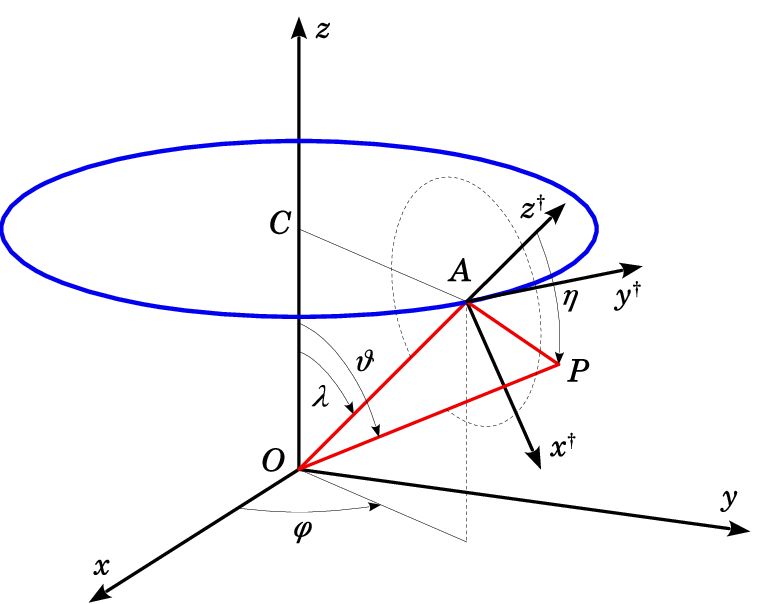}
\end{minipage}%
\begin{minipage}{0.35\textwidth}
	\begin{center}

	In the \textbf{S}$^{\dagger}$ frame: \\
 \vspace{0.2cm}

    $\zeta \equiv AP$ \\
	$k \equiv$ Normalized toroidal field strength \\
\vspace{0.2cm}
Magnetic field vector definition: 

\vspace{-0.5cm}
\begin{equation}
\mathbf{B} = \frac{1}{\sqrt{k^2 + \zeta^2}} \left (\zeta \cos\eta, k, -\zeta\sin\eta \right )
\label{eqn:magnetic_field_def}
\end{equation}

	\end{center}
\end{minipage}
\begin{minipage}[c]{0.3\textwidth}
	\includegraphics[width=\textwidth]{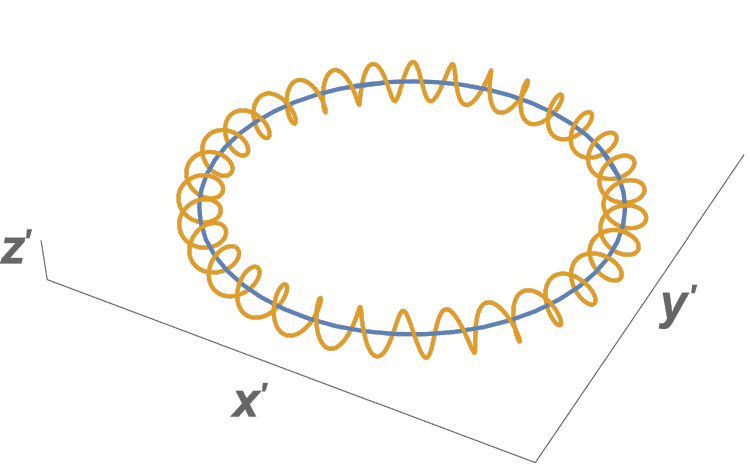}
\end{minipage}%
\caption{\textit{Left panel}: 
Magnetic flux rope geometry in the solar frame $\mathbf{S} \equiv (O; x, y, z)$ and
in the local frame $\mathbf{S}^{\dagger} \equiv (A; x^{\dagger}, y^{\dagger}, z^{\dagger})$. 
The origin $O$ of $\mathbf{S}$ corresponds to the Sun's center, and the $z$-axis to its 
rotation axis. The point $P$ is an arbitrary location in the flux rope domain, lying on the $ACO\equiv\langle z^\dagger,x^\dagger\rangle$ plane.
The axis of the flux rope is represented by the blue circle, tangent to the $y^\dagger$-axis in $A$.
\textit{Right panel:} Example of the magnetic field line of a helical flux rope with right-handed (RH) chirality, for a fixed distance \textbf{$AP$} from the flux-rope axis (colored in blue).}

\label{fig:geometry}
\end{figure*}

To compute the Stokes profiles emerging from a coronal plasma model,
we employed the 
\texttt{ScatPolSlab} code developed at the High Altitude
Observatory. 
\texttt{ScatPolSlab} computes the Stokes profiles of the radiation 
scattered by each voxel
of 
plasma in a given  stellar atmosphere, illuminated by a prescribed radiation
field at the lower boundary of the atmosphere.
The code can treat general magnetic regimes, ranging continuously from zero field 
strength to the complete Paschen-Back effect, and can accept any atomic structure
satisfying the Russell-Saunders (or $LS$) coupling scheme \cite[see, e.g.,][]{2004ASSL..307.....L}, including atoms with non-zero nuclear spin (hyperfine structure). The 
theoretical bases of \texttt{ScatPolSlab} are given in \cite{2005ApJ...624.1025C}.
The code adopts the formalism of the atomic density matrix to take into account the effects of 
quantum coherence among the atomic levels, responsible for the Hanle effect and for
polarization phenomena induced by level-crossing interference. It allows the user to specify
general illumination conditions, including surface inhomogeneities (e.g., sunspots and plage regions), and
it can model the polarization effects due to the Doppler dimming of the radiation 
propagating through the atmosphere, in the presence of plasma velocity fields. The code was originally written in \texttt{Fortan77}, but takes advantage of vector operations made available by \texttt{Fortran90} for both speed and readability.

In order to run \texttt{ScatPolSlab}, the user must specify the plasma and magnetic models in a 3D volume of the solar corona, sampled by a uniform Cartesian grid, and the radiation field at the lower boundary of the volume. At each grid point, the code solves for the density matrix of the atom in the single-scattering approximation,\footnote{In this approximation, the statistical equilibrium at any point in the corona is fully determined by the radiation field from the lower boundary reaching that point, possibly modified by the local plasma collisions. In other words, the coronal plasma is assumed to be optically thin towards the radiation source, even if it is allowed to be optically thick towards the observer.} and from it the absorption matrix and emissivity vector of the locally scattered radiation \cite[see][]{2005ApJ...624.1025C} are calculated at the same point. With this information available at every point in the coronal volume, the code then numerically solves the radiative transfer equation for polarized radiation along each of the LOS through the volume, 
 for all spectral lines 
of interest. 
The integration algorithm assumes that each grid point is representative of the 
magnetic and plasma properties of a uniform slab centered at that point, with a 
geometric thickness given by the grid step along the line of sight (LOS) \cite[see Sect.\ 8.3.a of][]{2004ASSL..307.....L}.

\texttt{ScatPolSlab} can also model resonance lines with partial redistribution, 
 for transitions in a 3-term atomic system of the $\Lambda$ type \citep{2016ApJ...824..135C}, although this feature is not utilized for the present work, as the radiation from the photosphere seen by the \ion{He}{1} atoms in the corona can  be assumed to be spectrally flat (regime of complete redistribution). In particular, for
our modeling effort, we adopted the 3-term \ion{He}{1} atomic model that includes the
transitions of the 1083\,nm and D$_3$ multiplets, and the chromospheric spectrum at those two wavelengths is practically featureless.

The plasma and magnetic models of the solar corona are specified in the solar reference frame $\mathbf{S}\equiv(O;x,y,z)$, where $O$ is the Sun's center, and the $z$ axis is the rotation axis. The observer's reference frame 
$\mathbf{S'}\equiv(O;x',y',z')$, used in the code to specify the Cartesian grid of the solar atmosphere, is such that the $x'$ axis is the LOS through disk center, whereas the choice of the orientation of the $y'$ and $z'$ axes on the POS is arbitrary. Standard choices have the $z'$ axis directed towards the ecliptic North, the geocentric North, or aligned with the projection of the Sun's rotation axis on the POS.
Because in our model we assume that the Sun's rotation axis lies on the POS, we simply choose $\mathbf{S'}\equiv\mathbf{S}$.
However, \texttt{ScatPolSlab} allows the user to choose any orientation of the solar frame $\mathbf{S}$ within the 3D mesh specified in 
the observer's frame $\mathbf{S}'$.
The code also takes into account the proper perspective correction of the $(y',z')$ coordinates
on the field of view (FOV) as a function of $x'$, due to the finite distance of the Sun from the observer.

The main purpose of this work is to investigate the polarization signatures of a helical magnetic flux rope 
suspended in the solar corona. We adopt an analytic magnetic model for a toroidal
helical flux rope, illustrated in the right panel
of Fig.~\ref{fig:geometry} and described by Eq.~(\ref{eqn:magnetic_field_def}) in the figure caption. This simplified 
model allows us to study the impact of different parameters of 
the helical flux rope on the \ion{He}{1} 1083 nm polarization signatures,
without the additional complexities of realistic 3D magnetohydrodynamic (MHD)
models. We postpone the study 
of such more realistic simulations to a future publication. 

We explore 
different orientations of the flux-rope axis 
(see Fig.~\ref{fig:geometry}, left panel), the
sign of the helicity of the magnetic flux rope, and density ratio 
between the flux-rope plasma and the surrounding corona. We set the helicity parameter such that the flux rope magnetic field has 32 coils, producing realistic twists, similar to those derived from
observations and extrapolations of CMEs \citep{Chen2017}. By choosing an integer number of 
revolutions of the magnetic field lines along the axis, we ensure the divergence free condition of the 
magnetic field (see Fig.~\ref{fig:geometry}).
For all runs in this work, we assumed a magnetic strength decreasing
radially away from the center of the flux rope with the square of the \emph{axial} distance, with a characteristic
scale length of 0.01\,$R_{\odot}$. 

For all calculations we assumed the background coronal plasma 
to be field free, with a number density of orthohelium of 1\,cm$^{-3}$
at the base of the corona, and an
exponential decay with 1\,$R_{\odot}$ scale height.
The denser region on the axis of the flux rope, mimicking a prominence condensation, is parameterized by a 
plasma density parameter $\rho_0$ with a value of 100\,cm$^{-3}$. The prominence plasma
density decreases
radially from the center of the flux rope with the square of the distance from the axis, with the same characteristic length scale as the magnetic field.
The aforementioned densities for the quiescent corona and the prominence condensation were computed using CHIANTI version 10 \citep{CHIANTI1,CHIANTI2}. We computed
those densities by estimating the ionization fraction of neutral helium in coronal and 
prominence conditions, and then estimated the populations of orthohelium in both
cases. The prominence density adopted in our computations agree with
previous prominence models, as well as it produces realistic emergent
intensity from our computations~\citep{Parenti2014}. The plasma temperature increases radially from the flux-rope axis 
with the same length scale as the magnetic field and the density. The plasma temperature
in our model is set to be 10$^4$\,K along the flux-rope axis, and it increases to a 
fixed value of 10$^6$\,K in the quiescent corona.

The mesh has 90 points along each LOS and 
180 points along each of the two spatial dimensions on the POS.
Numerical experiments 
with 
five times more points along the LOS showed very similar results, which
justified using only 90 points as a compromise 
between speed and accuracy in our numerical experiments. A typical run 
for a $90{\times}180{\times}180$ domain took about 26 CPU hours
on a Intel Xeon E5-2697V4 processor (hosted on the NSF NCAR Cheyenne supercomputer).

\section{Linear polarization of \ion{He}{1} 1083 as a coronal magnetic field diagnostic}
\label{sec:Results_observables}

To illustrate the diagnostic potential of \ion{He}{1} 1083 in the solar corona,
we provide various examples of measurable polarization signatures of 
coronal flux ropes observed from different viewpoints. We also 
explore the case of a global solar dipole field discussed in 
Section~\ref{subsec:Dipole}.
Despite the adoption of highly idealized models, the
results are indicative enough
for demonstrating the potential and inherent limitations of the Hanle-effect diagnostics of coronal fields.

First, at every point of the mesh, we define a local 
reference frame with its vertical axis
oriented along the local radial direction through Sun's center.
In this reference frame, we assume that the incoming radiation is 
axially symmetric around the 
radial direction. This is a good approximation for a 
flux rope overlying quiet-Sun regions, 
at a large enough height that surface inhomogeneities at the granulation scale
($\sim 1/2\arcsec$) average out within the radiation cone seen by a scattering point in the flux rope.
Under this approximation, the radiation field at the scattering point is fully specified 
by the radiation temperature of the Sun at the wavelength of the atomic transition, 
the center-to-limb variation (CLV) of the surface brightness at that wavelength, 
and the height of the scattering point above the surface. As we pointed out earlier, the solar radiation around 1083\,nm is sufficiently devoid of spectral features that the illumination of the flux rope can be considered spectrally flat, thus validating the approximation of complete redistribution adopted in our model.

In the absence of a magnetic field, or when the field is purely radial, the direction of the linear polarization of \ion{He}{1} 1083 observed at a point in the FOV will be 
perpendicular to the projected solar radius through the point.
Thus, if we adopt this direction as the local reference of polarization at each
point in the FOV, a linear polarization map of the Sun would show a radially dependent Stokes $Q$ and zero Stokes $U$ everywhere, in this special case. 
In contrast, a non-zero Stokes $U$ signal at any point in the FOV
is a direct indication of the presence of a non-radial magnetic field somewhere along the LOS through that point.
For this reason, we adopt such a polarization
reference frame for all polarization maps shown in this work.

\subsection{Global dipole field of the Sun}
\label{subsec:Dipole}

\begin{figure}[!ht]
	\centering
	\includegraphics[width=0.49\textwidth]{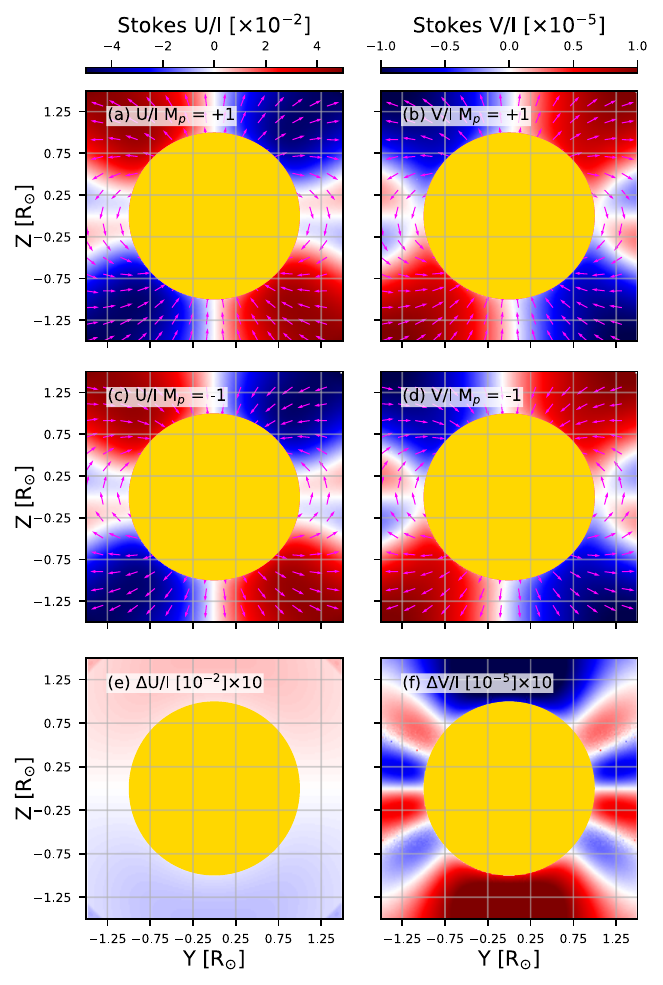}
	\caption{\ion{He}{1} 1083 wavelength-integrated
     Stokes $U$/$I$ (left column) and $V$/$I$ (right column) produced by a 
     global dipolar magnetic field. The top row corresponds to 
    a dipole field with positive polarity at the North solar pole. 
    The middle row shows the same dipole field with inverted polarity. The magenta arrows in those plots trace the magnetic field lines in the POS for the two models. 
	The bottom row shows the polarization difference between the two models. While in the pure Zeeman case this difference would be exactly zero, in the case of the Hanle effect there are subtle differences due to the symmetry breaking of non-isotropic scattering. As expected,
	the amplitudes of the Hanle-induced Stokes $U$ polarization are significantly
    larger than those of the Zeeman-dominated Stokes $V$ polarization.}
	\label{fig:dipole_field}
\end{figure}

\begin{figure*}[t!]
    \centering
    \includegraphics[width=\textwidth]{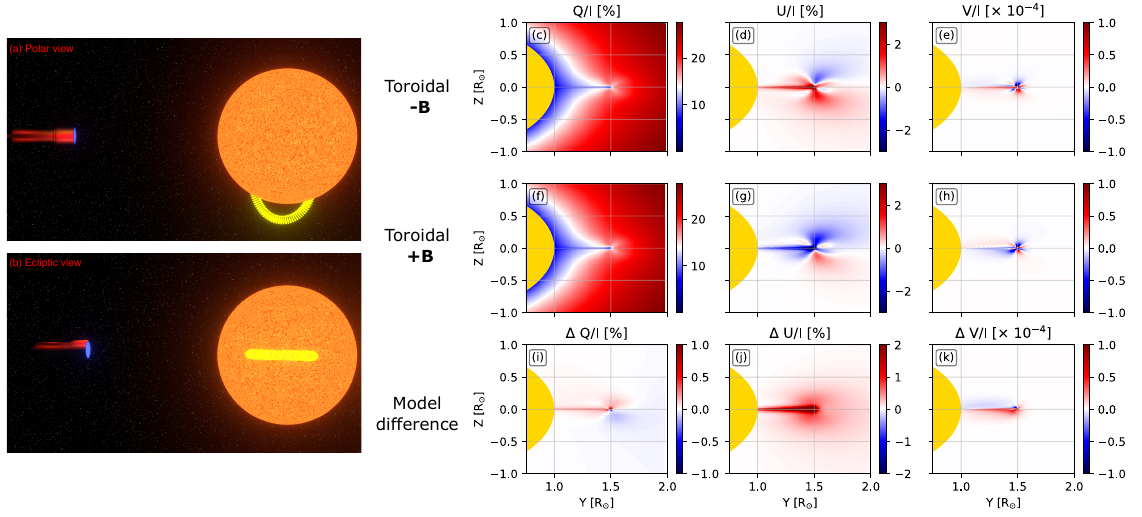}
    \caption{Model of a helical flux rope laying in the ecliptic plane, and intersecting the YZ plane 
    at X = $0\,R_{\odot}$ and Y = $1.5\,R_{\odot}$. The flux rope has a radius R = $0.5\,R_{\odot}$ and it is anchored to the Sun.
    Panels (a) and (b) illustrate the observing 
    geometry (in this case, a ``quadrature'' configuration), where
    the telescope symbol represents the observer. The 
    spectrally integrated 
    Stokes signals ($Q,U,V$) are
    shown in panels (c) through (h) for the longitudinal field
    of the flux rope pointing either towards (-\,$\mathbf{B}$) or away from (+\,$\mathbf{B}$) the observer (cf.~Fig.~\ref{fig:geometry}). Panels (i) through (k)
    show
    the corresponding differences between the two cases. 
    We adopt the same convention for the local reference direction of linear polarization as described in 
    Figure~\ref{fig:dipole_field}. 
    }
    \label{fig:ScatPolSlab_result_1}
\end{figure*}

To examine the potential of the Hanle effect as a
magnetic field diagnostic of the solar corona, we computed the polarization signature
of the \ion{He}{1} 1083 multiplet, assuming a global dipolar field with a surface
strength of 1 gauss at the solar equator. Similarly to the work of
\citet{2012ApJ...760....7M} on the modeling of Hanle
signatures from stellar surface magnetic fields, we compute the 
polarization signals
originating at coronal heights.

The maps of Stokes U and V, wavelength-integrated across the line profiles and normalized by the integrated intensity, are shown in Figure~\ref{fig:dipole_field}, respectively in the left and right columns.
The first two rows show the Stokes maps for two opposite polarities of the same magnetic dipole field. The last row
(panels (e) and (f)) shows the polarization difference between the two models (scaled tenfold 
to show the differences on the same colorbar).
The most striking feature of Figure~\ref{fig:dipole_field} is the apparent 
similarity of the global polarization signatures for the two dipole-field polarities, with only a difference
of the order of 0.1\% (see the last row of Figure~\ref{fig:dipole_field}). 
This is mainly due to the very low opacity of the coronal plasma for the assumed \ion{He}{1} density distribution,
combined with the Hanle polarization symmetries of a non-tilted dipole field. These symmetries make so that, by inverting the polarity of the dipole field, the back and front hemispheres exchange roles in producing the same polarization pattern. Therefore, in an optically thin corona, the sum of the contributions from the two hemispheres for an off-limb LOS will be practically identical for the two polarities of the dipole field.

\begin{figure*}[t!]
    \centering
    \includegraphics[width=\textwidth]{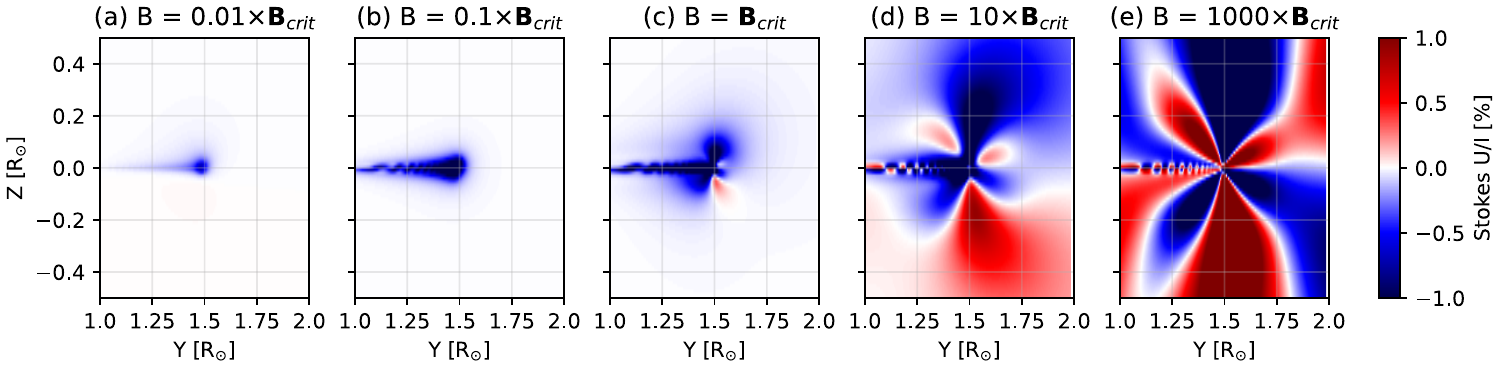}
    \caption{Changing the magnetic field strength creates different polarization
    Stokes $U$ signals, but the overall polarization sign remains the same in the 
    unsaturated Hanle regime. We present 
    the same geometrical setting as in Figure~\ref{fig:ScatPolSlab_result_1}
    with positive toroidal field component but for varying axial field strengths. For panels (a)-(e) the field strength
    is changing from 0.01 of the Hanle saturation strength (for the \ion{He}{1} 1083 
    $B_{\rm crit} \approx 0.7$\,G) to 1000 times the saturation strength.
    }
    \label{fig:Bfield_variation}
\end{figure*}

 \begin{figure*}[ht!]
 	\includegraphics[width=\textwidth]{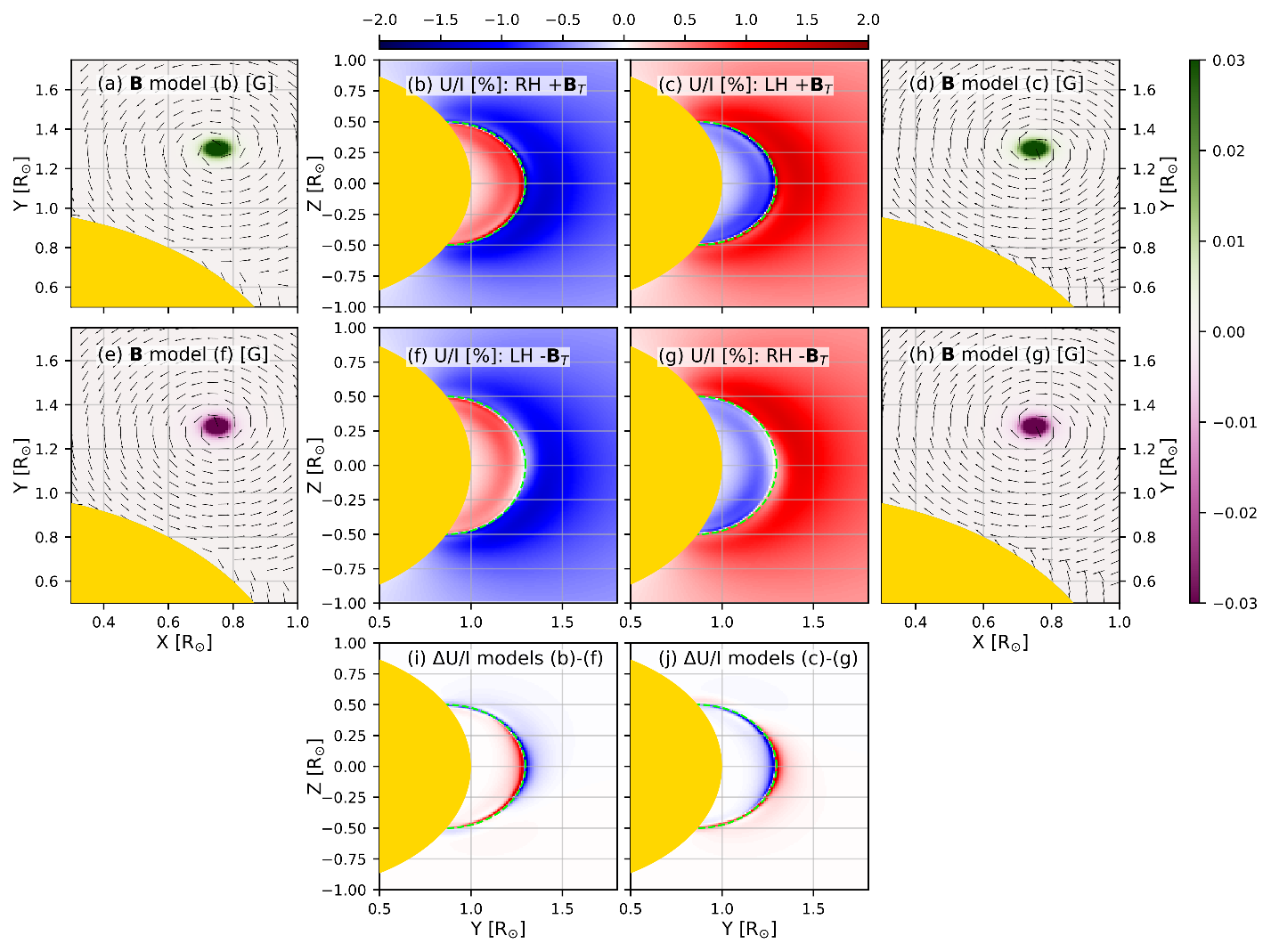}
 	\caption{Magnetic flux rope perpendicular to the ecliptic plane observed at 
    $60^\circ$ of heliocentric longitude. The flux rope is anchored to the Sun
    as in Figure~\ref{fig:ScatPolSlab_result_1} and intersects the 
    ecliptic at $X = 0.75\,R_\odot$, $Y = 1.3\,R_\odot$.
    Panels (a), (d), (e), (h) show slices of the 
    magnetic field in the ecliptic plane; the observer 
 	is to the right, at 1 AU ($X=214~R_\odot$);
    Panels (b), (c), (f), (g) show the Stokes $U/I$ maps
    for the different magnetic configurations, distinguished by their chirality (right handed, RH; left handed, LH), and orientation of the axial field (northward,
    $+B_{T}$; southward, $-B_{T}$).
    	Note that the globalpattern of Stokes $U$ is determined by the poloidal component of the field, rather than by
    the axial component as in the case of the edge-on flux rope. 
    The differences between the pairs of resembling Stokes $U$/$I$
    maps are on the order of a few tenths of a percent,
    as shown in panels (i) and (j). The dashed lime line outlines the core of 
    the flux rope.}
 	\label{fig:L45_perp_FR}
 \end{figure*}
 
\begin{figure*}[ht!]
	\includegraphics[width=\textwidth]{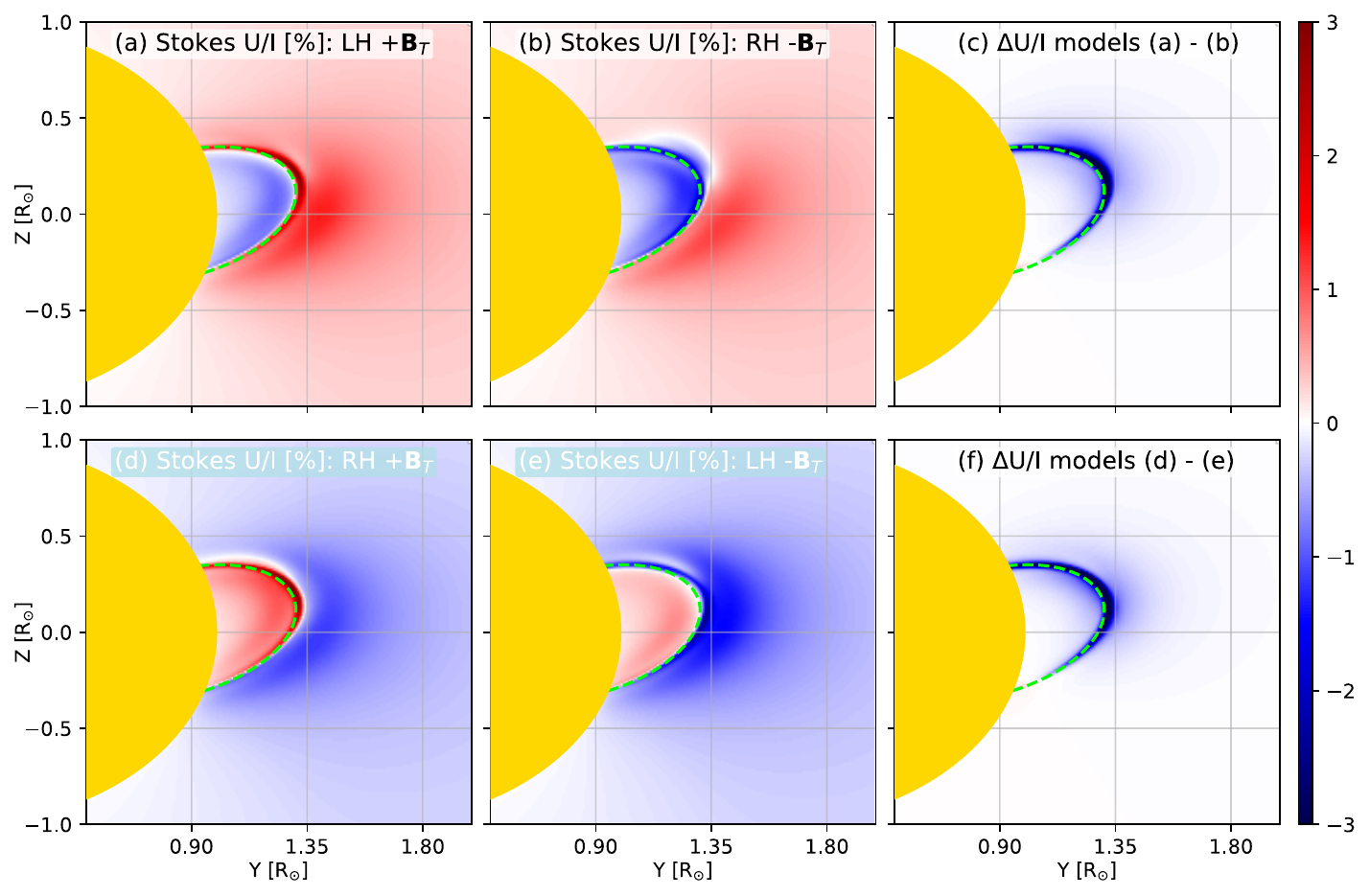}
	\caption{Similar magnetic flux-rope setup to the one described in 
	Figure~\ref{fig:L45_perp_FR}, but with the flux rope tilted at 45$^\circ$ to the ecliptic
	    plane. Again, the poloidal magnetic field direction determines the global Stokes $U$
	    pattern, as shown in panels (a), (b), (c), and (d). The differences between the models, 
	    shown in panel (c) for models (a)-(b) and panel (f) for models (d)-(e) 
        illustrate that there are
	    detectable Stokes $U$ differences on the order of 1\%. The lime green 
        dotted line follows the core of the flux rope.}
	\label{fig:L45_tilted_FR}
\end{figure*}

We must note that, for a static atmosphere, the Stokes V difference map between the two dipole configurations would be zero in the pure Zeeman case, even in an optically thick corona, because of the intrinsic symmetries of the Zeeman effect. On the other hand, when the atomic level interference of the fine structure of orthohelium is taken into account, a conversion mechanism between 
atomic alignment and atomic orientation sets in \citep{1984ApJ...278..863K}, which becomes the main source of the
net circular polarization emerging in our models, as seen in panels 
(b) and (d) \citep[for a detailed description of this effect, and of its application to the \ion{He}{1} chromospheric lines, see][]{2004ASSL..307.....L,2008pps..book..247C}. 

Therefore, 
in order to utilize the Hanle effect as a magnetic diagnostic of the corona, even in the 
unsaturated regime, 
we need a certain amount of asymmetry in the way that the magnetic field and/or the 
scattered radiation are distributed along the LOS, 
so to be able to discern the orientation of the magnetic field.
The case of an expanding flux rope, or
 other localized magnetic structure in the global solar corona, are clearly ideal candidates
 for the application of this diagnostic \cite[see also][]{2019ApJ...883...55Z}.
In the following subsections, we present examples of the polarization signatures of
\ion{He}{1} 1083 for
models of flux ropes of increasing geometric complexity. 

\subsection{Flux rope in the ecliptic plane observed at quadrature}
\label{subsec:edgeon_FR}

The simplest observing scenario is that of a  helical coronal
flux rope lying in the ecliptic and observed in quadrature. We show the 
orientation of such a flux rope in panels (a) and 
(b) of Figure~\ref{fig:ScatPolSlab_result_1}; the flux rope is represented by the bright
``slinky'' structure, while
the observer is indicated by the telescope symbol. The flux rope has a 
radius of $0.5\,R_{\odot}$ and its center is situated 
in the ecliptic plane at quadrature with the observer. The axial field strength, which 
corresponds to the field directed along the LOS where the flux-rope axis intersects the 
POS, is set to 1\,G. We computed the 
\ion{He}{1} 1083 polarization when the axial field 
points towards or away from the observer, the results being shown in 
the first and second row of Figure~\ref{fig:ScatPolSlab_result_1}, respectively (panels (c)-(h)).

As expected, the two cases of opposite axial magnetic 
field show opposite signs of the Hanle-induced Stokes $U$ polarization (panels (d) and (g)). In particular, this implies that the magnetic-induced rotation of the linear polarization with respect to the local tangent to the limb is of opposite sign in the two cases.

Notably, the expected Stokes $U$ amplitude is of the order of 2\% of the 
line intensity, hence, readily accessible to an instrument with easily attainable polarimetric sensitivity. The 
signal also shows spatially coherent large-scale structure, over which one could potentially bin the signals 
in order to further improve the sensitivity of the measurement. 
In contrast, the Stokes $V/I$ signal is expected to be 
at most around $10^{-4}$ (panels (e) and (h)),
hence quite difficult to detect
without the use of large instruments, similarly to the case of the Zeeman-based magnetic diagnostics of the solar corona \citep{2023BAAS...55c.392T}. 

For this type of flux-rope geometry, the Hanle-generated Stokes $U$ polarization turns out to be a robust diagnostic of the orientation of the axial field. To show this, we computed different FOV maps of Stokes $U$ for a range of axial field strengths spanning the full range of criticality of the Hanle effect. The results of Figure~\ref{fig:Bfield_variation} show how the polarization amplitude remains predominantly of the same sign throughout the unsaturated Hanle 
regime ($B \lesssim 10\,B_{\rm crit}$), determined by the orientation of the axial field. In the saturation limit ($B\gg  10\,B_{\rm crit}$), the polarization pattern becomes instead closely correlated with the direction of the field projection on the POS. At the same time we also note the appearance of the characteristic null-polarization surfaces due to the Van Vleck effect \cite[e.g.,][]{2002ApJ...568.1056C}. This demonstrates that the Stokes $U$ polarization is a robust tracer of the orientation of the axial field over 
the range of magnetic field strengths of the unsaturated Hanle effect.

\subsection{Flux rope outside of the ecliptic plane: The case for L4/L5 
observations}
\label{subsec:Results_mag_field_L4_L5}

When the scattering angle of the radiation is close to $90^\circ$ (near limb) or $0^\circ$ (near disk center), the Hanle-effect linear polarization suffers from the same $180^\circ$ ambiguity characteristic of the Zeeman effect. This is a fundamental obstacle to the discernment of the orientation of the field via linear polarization measurements. Unlike the Zeeman effect, this ambiguity is lifted for scattering angles significantly different from those degenerate configurations \cite[e.g.,][]{1993ApJ...411L..49L,2002ApJ...568.1056C}. 
For this reason, in practical applications aiming at the determination of $B_z$, it is critical to target such unambiguous scattering configurations.

To demonstrate this point, in this section we explore the case of a flux rope laying orthogonally to the ecliptic and observed
at a heliographic longitude of $60^\circ$. This scenario mimics what
an observer at the Lagrangian points L4 or L5 would see in
the case of a prominence eruption directed towards Earth. Other than the orientation of the flux rope, all 
other geometric and magnetic characteristics are identical to the case discussed in
Sect.~\ref{subsec:edgeon_FR}.

The Stokes $U$ maps for this case are shown 
in panels (b), (c), (f), and (g) of Figure~\ref{fig:L45_perp_FR}, where we considered all possible four
combinations of different chirality and axial-field orientations.
The labels in those panels stand 
for left/right-handed (LH/RH) flux-rope magnetic field, while the sign of the 
axial field $B_T$ distinguishes the orientation of the magnetic field 
in the ecliptic. Positive/negative value of $B_T$ point toward the north/south  pole of the ecliptic, as shown in the magnetic-field cuts in the 
ecliptic plane of panels (a), (d), (e), and (h).
In these panels, the color of the 
magnetic map distinguishes the out-of-plane (axial) field orientation
(green for positive, or northward; violet for negative, or southward).
 
The large-scale patterns of Stokes $U/I$ appear to be qualitatively similar for configurations for which the orientation of the flux-rope \emph{poloidal} field (i.e., the field component lying on the $ACO$ plane of Figure~\ref{fig:geometry}) is the same, regardless of the orientation of the axial field. This is to be expected, as the $U$ signal from the plasma surrounding the flux rope is largely dominated by the sign of the field projection along the LOS.
However, quantitatively the results are not identical, as shown by the difference panels (i) and (j), and are at a level of a few tenths of a percent, hence potentially within reach of an instrument with a polarimetric accuracy of $10^{-3}$ or better.
It is important to remark that these differences are real, and not a numerical artifact of the discretization of the domain, as the structures in these maps are fully resolved at the density level of our mesh.
Such differences are concentrated in the vicinity of
the flux-rope core, but need to be discerned over the nearly tenfold larger polarization amplitudes characterizing the large-scale patterns seen in the Stokes $U/I$ maps, which are comparable
with the amplitudes observed in the case of the edge-on flux rope of the previous section.

We also computed 
polarization maps for the same flux rope model, after tilting it by 
$45^\circ$ on the ecliptic. The results are shown in Figure~\ref{fig:L45_tilted_FR}, 
for the same combinations of chirality and 
axial-field orientations as in Figure~\ref{fig:L45_perp_FR}. The differences between configurations 
that lead to similar large-scale polarization patterns are 
shown in panels (c) and (f). Note that in this case the polarization variations that are localized around the flux rope
are much larger than in the former case, now being of the order 
of more than 3\%. As predictable, when the axial field
of the flux rope has a non-null projection onto the LOS as it happens with 
such a tilted flux-rope model, different
axial-field orientations produce Stokes $U/I$ signatures of opposite signs, similarly to the cases considered for Figure~\ref{fig:ScatPolSlab_result_1}.
This is again clearly seen in the difference maps (panels (c) and (f)).

 \begin{figure*}[t!]
 	\includegraphics[width=\textwidth]{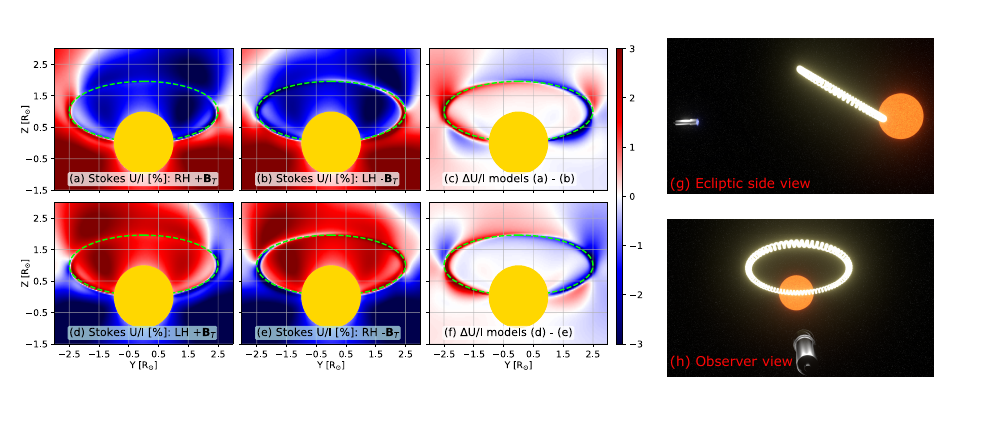}
 	\caption{\ion{He}{1} 1083 Stokes $U$/$I$ synthetic observables 
     from a halo CME-like flux rope oriented toward the 
     observer. Panels (a)-(f) follow the notation of Figure~\ref{fig:L45_tilted_FR}.
     Panels (g) and (h) show the orientation of the flux rope in respect to the 
     observer, who is marked as the telescope. The large scale degeneracy
     of the opposite chirality and axial field combinations, previously noted in 
     Figure~\ref{fig:L45_perp_FR}, persists in this case as shown in panels (a)-(b) and 
     (d)-(e). The green dashed line indicates the POS projection of the 
     flux rope core.}
 	\label{fig:Halo_CME}
 \end{figure*}

The previous examples show that polarimetry at the $10^{-3}$ sensitivity level is necessary in order to distinguish among the possible different combinations of chirality and axial-field orientations of the flux rope, which give rise to distinguishable polarization signatures as shown in the Stokes $U/I$ maps of Figures  \ref{fig:L45_perp_FR} and \ref{fig:L45_tilted_FR} (see also Figure~\ref{fig:L45_noise} below).

Additional observational constraints can further help reducing ambiguities in the inference of the magnetic topology of the flux-rope, or may even be required in the case of observations with polarimetric noise larger than $10^{-3}$ or with low spatial resolution. For example, the axial-field orientation can be inferred through
magnetic-field extrapolations based on photospheric magnetic 
fields \citep[see, e.g.,][]{Wiegelmann2012}. On the other hand, observations constraining the flux-rope chirality can be conducted during its quiescent evolution, as chirality is known to be a largely preserved property of a flux rope \citep{1994GeoRL..21..241R}. A popular method relies on observations of the direction of the prominence barbs in the  hydrogen H-$\alpha$ line at 656.3 nm \citep{1998ASPC..150..419M, 2008SoPh..250...31M}.

\subsection{Observing halo CMEs}
\label{sec:Results_mag_field_halo}

Halo CMEs are observed in coronographic observations 
whenever the path of the CME is approximately aligned with the LOS 
\citep{1982ApJ...263L.101H}. When this coincides with the Sun-Earth line,
such a class of CMEs poses a 
significant threat due to their high geo-effectiveness and commonly delayed detection   \citep{2000JGR...105.7491W, 2006SpWea...410003M}. 
In order to investigate the potential of \ion{He}{1} 1083 diagnostic for determining the 
magnetic field topology (mainly the $B_z$ direction) of halo CMEs, we modeled different geometries of the underlying flux rope, and viewpoints other than
the Sun-Earth line.
Because halo CMEs can only be detected when the linear expansion of the ejected 
material has well exceeded the diameter of the solar disk, we considered a flux rope 
with a radius of $2.5~R_\odot$, and its center located 
on the LOS at a projected height of $1~R_\odot$ above the surface (i.e., at 
coordinates $(Y,Z)=(0,2)~R_\odot$ with an inclination angle to the ecliptic
of $24^\circ$)
and axial field of 1 gauss. The flux rope may still be anchored to the solar 
surface, but the leading edge of the expansion will project outside of the solar disk, 
and therefore observable with coronagraphic instruments.

Figure~\ref{fig:Halo_CME} shows the results of the halo CME computation. 
Despite the prevalent forward-scattering geometry characterizing this example --- which is expected to produce lower scattering polarization signals than the previously discussed examples,  we still predict a significant
Stokes $U$ amplitude, for the assumed strength of the axial field, on the order of a few percent of the line intensity.The 
degeneracy seen between the LH and RH chirality cases with opposite axial fields 
described in Section~\ref{subsec:Results_mag_field_L4_L5} persists 
for the general Stokes $U$ pattern.
However, we note the large 
differences in $U/I$ (of the order of a few percent) in the far edges of the
halo CME seen in Figure~\ref{fig:Halo_CME} (around coordinates $(Y,Z) = (\pm 2.5,0.5)\; R_{\odot}$) which are relevant for 
the disambiguation between the different topologies of geo-effective CMEs. Similarly
to the result for the 45$^{\circ}$ tilted flux rope to the ecliptic in 
Figure~\ref{fig:L45_tilted_FR}, those regions of strong longitudinal field 
could be used for distinguishing between the different magnetic geometries.

\subsection{Effects of the density on the observed polarization}
\label{subsec:rho_dependence}

\begin{figure*}[ht!]
\includegraphics[width=\textwidth]{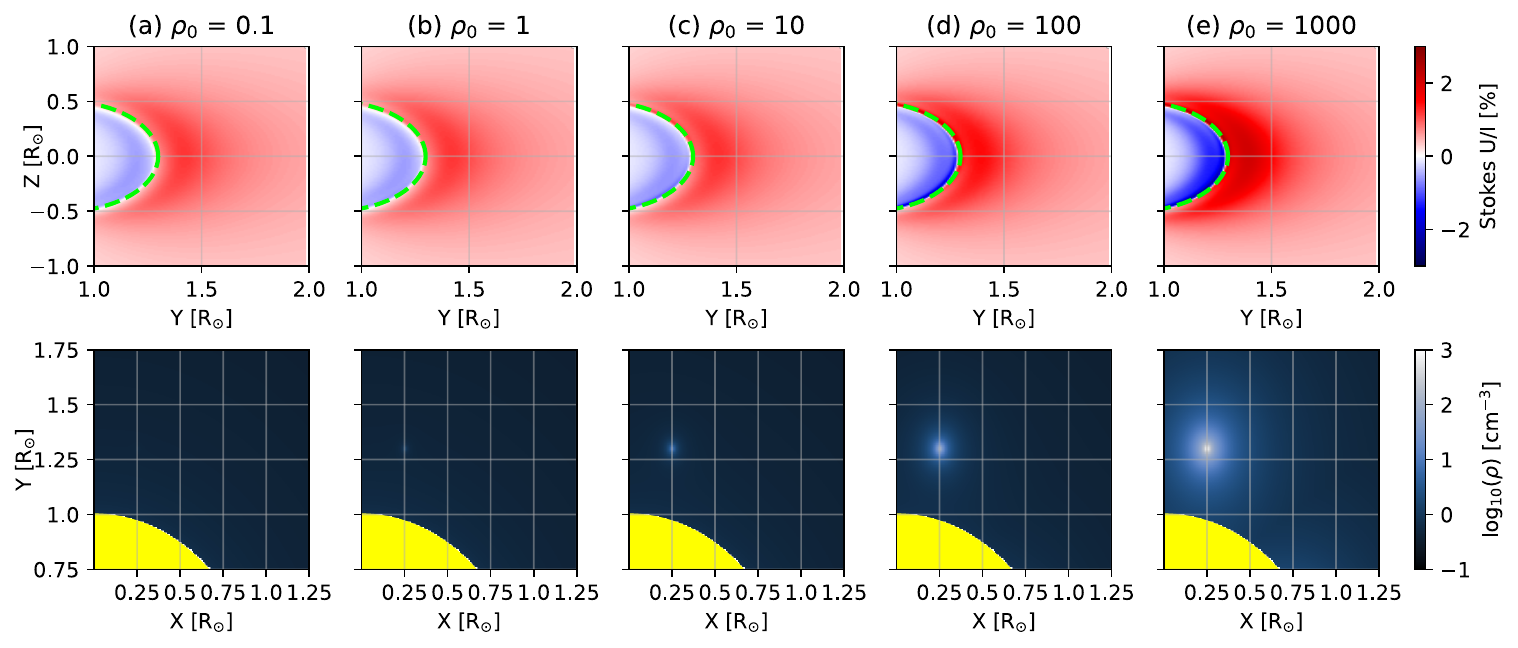}
	\caption{\emph{Top Row:} \ion{He}{1} 1083 Stokes $U$/$I$ maps for the
    flux-rope model of Figure~\ref{fig:L45_perp_FR}
    panel (c), with different central density $\rho_0$
    of the flux rope. The values of $\rho_0$ 
    in units of cm$^{-3}$ are annotated at the top of the panels. The lime green
    dashed line shows the core of the flux rope.
    \emph{Bottom row}: Cross section on the $Z=0$ (ecliptic) plane of the number density 
    of scattering atoms in the solar corona for the corresponding models in the top row. }
	\label{fig:rho_dependence}
\end{figure*}

Our atmospheric model has a simplified structure, with multiple parameters 
being prescribed manually with values inspired by real observations
\citep{2012LRSP....9....3W}.
The parameter with the largest effect on our computations is the 
density difference between the prominence-like structure
associated with the model flux rope and the density 
of the surrounding corona.
The density in our model is prescribed by
the sum of the background coronal
density profile 
$\rho_{\rm atm}(h)$, exponentially decreasing with height,
and the density $\rho_{0}$ of the magnetic flux rope parameterized through the
magnetic field strength. This additional density
$\rho_{0}$ of the magnetic 
structure is included to model the prominence condensation.
The total density of helium atoms determines the emissivity of the coronal plasma,
but the distribution along the LOS determines the polarization
signature emerging from the different components of the corona, properly taking into 
account the polarized emissivity and absorptivity of each voxel of plasma along the LOS.
To capture the effects of different density distributions, we ran a grid 
of models, similar to the one presented in
Figure~\ref{fig:L45_perp_FR} panel (b),
where we changed the parameter $\rho_0$ between 0.1 to 10$^3$\,cm$^{-3}$. 
The results of these calculations are shown in 
Figure~\ref{fig:rho_dependence}.

For the case of modest to no density enhancement in the flux rope, panels (a)-(c) in Figure~\ref{fig:rho_dependence}, we see about the same signals 
as described before. This corroborates the robustness of the proposed method 
over a range of 2 orders of magnitude of plasma densities. The results do 
not change significantly until the overdensity 
in the flux rope becomes significant, as shown in cases (d) and (e) with 
$\rho_0\gtrsim 10^2~\rm cm^{-3}$. In this case the total Stokes $U$ polarization 
increases as the emitting material in the core of the flux rope becomes
 optically thicker, and samples the plasma closer to the core of the flux rope. 

\section{Discussion}
\label{sec:Discussion}

 \begin{figure*}
 	\includegraphics[width=\textwidth]{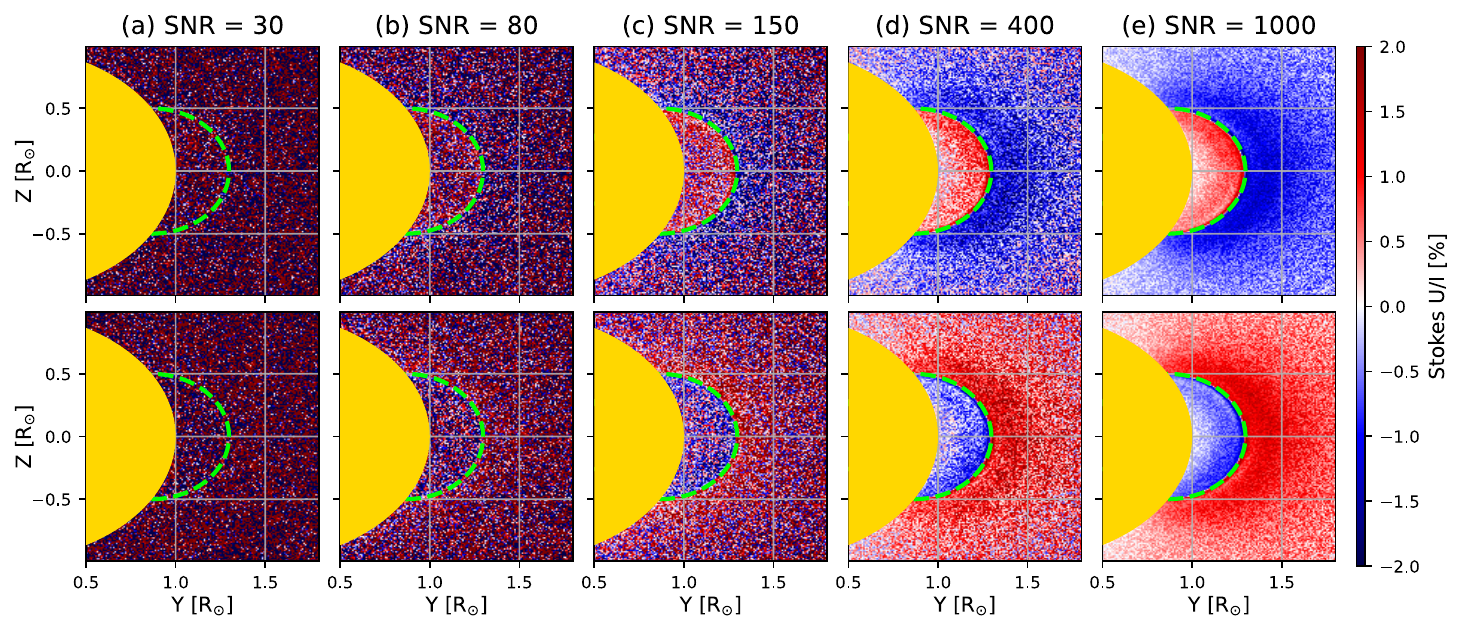}
 	\caption{Impact of noise on the Stokes $U$/$I$ synthetic observations of a flux rope
             on the Sun-Earth line observed from L4/L5. This figure shows the effect of
             decreasing noise levels for the model in Figure~\ref{fig:L45_perp_FR} panel 
             (b) in the top row and in the second row for the model in 
             Figure~\ref{fig:L45_perp_FR} panel (c). The SNR listed 
             in the panel titles. For the instrument proposed in the text we expect the 
             scenario of SNR\,$\sim$\,1000 along the core of the flux rope, corresponding 
             to column (e). The green dashed line shows the core of the flux rope.}
 	\label{fig:L45_noise}
 \end{figure*}

Most geo-effective CMEs originate from solar active regions
close to the Sun-Earth line or slightly toward the solar east limb, where the 
solar rotation contributes to the ballistic propagation of the CMEs towards 
Earth \citep{2021FrASS...8...79B}. 
The flux rope geometries described in this paper
are readily applicable to the majority of geoeffective CMEs. 
Establishing space-based polarimetric coronagraphic observations of \ion{He}{1} 1083 for space-weather forecast, specifically at L4/L5, would enable direct observations of 
the Sun-Earth line from a favorable viewpoint for 
Hanle magnetometry.
For this reason, we investigated the 
scenario of a flux rope oriented perpendicularly to the ecliptic
observed from the vantage point of a L4/L5 observatory -- i.e., approximately 
$\pm60^\circ$ away from the Sun-Earth line in heliocentric longitude. 
The \ion{He}{1} 1083 magnetometry approach works well for this case.
In reality, the most plausible scenario for an ICME is an intermediate one
between the one described in this section and the one studied in
Section~\ref{subsec:Results_mag_field_L4_L5}. Our study of the tilted 
flux rope, reported in Figure~\ref{fig:L45_tilted_FR}, clearly illustrates
that the Hanle effect of the \ion{He}{1} 1083 line represents
 a robust magnetic diagnostic for a realistic 
scenario of a flux rope tilted on the ecliptic. However, to fully exploit the
potential of this approach, additional observations, such as photospheric 
magnetic field estimates and H$\alpha$ filtergrams, could provide critical 
constraints on the geometry  
of the erupting flux rope, such as to break the degeneracy 
between the different possible combinations of chirality and 
axial field orientations, as illustrated in
Figures~\ref{fig:L45_perp_FR}, \ref{fig:L45_tilted_FR},
and \ref{fig:Halo_CME}.

An important consideration for the feasibility
of the proposed magnetometry approach is the signal-to-noise 
ratio (SNR) of potential coronal \ion{He}{1} 1083 observations, 
which we estimated with the following formula:
\begin{equation}
\mathrm{SNR} = \epsilon_{Q,U} \sqrt{\frac{I_{1083} \; 
A_{\rm tel} \; A_{\rm px} \; t_{\rm int} \;  T_{\rm tel} \; QE }{E_{\rm ph}}}
\label{eqn:SNR}    
\end{equation}
$\epsilon_{Q,U}$ is the modulation efficiency for the $Q$ and $U$ linear polarization (assumed to be identical); $I_{1083}$ is the wavelength-integrated specific intensity of the \ion{He}{1} 1083
line; $A_{\rm tel}$ is the area of the telescope aperture;  $A_{\rm px}$ is the angular area subtended 
by a pixel;  $t_{\rm int}$ is the integration time; $T_{\rm tel}$ is the transmissivity 
of the optical system; $QE$ is the quantum efficiency of the detector; and $E_{\rm ph}$ is the energy of a 1083\,nm photon.
We calculated the photon budget of a space-borne
imaging coronagraph and polarimeter, consisting of a telescope with a 5\,cm aperture
diameter
and 3.5\arcsec\ pixel, assuming a 0.5\,s integration time.
We estimated the throughput of such an instrument,
assuming: a primary objective and a field lens (0.9 transmission each); 
a heat rejection filter plus a narrow prefilter (0.5 total transmission);
a polarizing beam-splitter (0.45 transmission in each of the two channels);
a near-IR camera (0.9 QE); a throughput loss from vignetting of about 20\% (i.e., 0.8). 
By multiplying these transmissivity values for the individual elements, we obtain an 
effective throughput for such an instrument of about 0.13.

The proposed instrument observes the wavelength integrated intensity across
the \ion{He}{1} 1083 multiplet. 
Considering the average brightness in \ion{He}{1} 1083 of a typical ``hedgerow'' 
prominence (between $\sim 1/2\mathrm{-}1/50\,B_{\odot}$), based
on recent observations \citep{2015ApJ...802....3M}, we assume a very 
conservative value of 10$^{-2}$ $B_{\odot}$ for our model demonstration.
We note that these brightness estimates 
agree with the numerical work by \citet{2020ApJ...898...72D}. Our
model provides a SNR of about 1000 for the $Q$ and $U$ signals along the brighter 
core of the flux rope (case (e) in Figure~\ref{fig:L45_noise}), assuming
the observing conditions stated above, and an optimal 
modulation efficiency $\epsilon_{Q,U} = 1/\sqrt2$.
In Figure~\ref{fig:L45_noise}, we consider again the synthetic observations of 
\ion{He}{1} 1083 Stokes $U/I$ for the case of the L4/L5 flux rope model presented in 
Fig.~\ref{fig:L45_perp_FR},
and plot them for different SNR levels close to the limb, 
where for each pixel we have scaled the local SNR by the local brightness.
Due to the large scale pattern and large Stokes $U/I$ signal from
the Hanle effect, the different cases could be distinguishable even for very 
low SNR targets, such as the ones illustrated in the left columns having lower 
SNR.
Hence, even if the 
\ion{He}{1} 1083 brightness in the corona is an order of magnitude lower, or a
significant source of scattered light is introduced in the hypothetical 
instrument, by increasing the pixel size to 10\arcsec\ and the exposure time to 2 seconds, we would be able to achieve similar results to panel (e) of Figure~\ref{fig:L45_noise}.
This is a strong argument for pursuing the \ion{He}{1} 1083 coronal 
magnetometry approach discussed in this work. 

The most significant limitation of our study is the simplicity of 
the coronal and flux-rope models adopted,
as they lack a self-consistent magnetic and
thermodynamic treatment of the active region of origin, including
the prominence condensation. As shown in Section~\ref{subsec:rho_dependence}, 
there is a strong dependence of the observed polarization on the density distribution 
of the underlying coronal model. Other effects that were not considered in
this study are the
variations of both the magnetic field and the plasma density in the observed structures. 
These will be addressed in a future study, where we plan to reproduce the numerical 
experiments presented in this work using more realistic and self-consistent 3D radiative 
MHD models of prominence eruption from \cite{2018ApJ...862...54F}.

\section{Conclusions}
\label{sec:Conclusions}

This work shows how the magnetic field of flux ropes in the solar 
corona produces detectable signatures in the linear polarization of the \ion{He}{1} 1083 
nm atomic line through the Hanle effect, which can be used to infer the magnetic 
topology of these coronal structures, with a particular emphasis on the problem of 
discriminating the $B_z$ component of CMEs.
In particular, our
computations clearly present that diagnostics built on the wavelength-integrated 
polarization signals of
the \ion{He}{1} 1083 multiplet can rely on measurable signals of the order
of 10$^{-2}$ in the magnetic-induced Stokes $U/I$ polarization.

In the favorable case of a helical flux rope seen edge-on and
in quadrature above the solar limb, as shown in Figure~\ref{fig:ScatPolSlab_result_1},
a quantification of the axial magnetic field, which is necessary to estimate 
the magnetic energy stored in the flux rope, is readily provided
by the magnitude and sign of the wavelength-integrated Stokes $U$.

Another relevant magnetic geometry for space-weather 
applications corresponds to a flux rope
on the Sun-Earth line viewed from the L4/L5 vantage points. 
Our modeling examples in
Section~\ref{subsec:Results_mag_field_L4_L5} show that the different
combinations of chirality and orientation of the axial magnetic field produce distinguishable
polarization signatures, as illustrated in Figure~\ref{fig:L45_perp_FR}. However, we found the observable characteristics to be almost degenerate between pairs of opposite chirality and axial field directions, as described in Section~\ref{subsec:Results_mag_field_L4_L5} 
for flux ropes perpendicular to the ecliptic.
It is found that a polarimetric precision of $10^{-3}$ is required to distinguish 
between the different cases, ordinarily attainable with present day's instrumentation, 
whereas additional topological information readily available from photospheric and 
chromospheric data could further help constrain the magnetic field direction. 
In the case of a flux rope inclined on the ecliptic 
outside of the plane of the sky as shown in Figure~\ref{fig:L45_tilted_FR} this 
degeneracy is broken and the Hanle effect provides a robust magnetic geometry diagnostic.
Since the polarimetric observables depend on the orientation of the flux rope,
some modeling of the active region to infer its geometry would be required,
such as methods based on magnetic field extrapolations from photospheric data
\citep{Wiegelmann2012}. 

We also showed that halo CMEs produce Hanle-induced Stokes $U$ signal in the \ion{He}{1}
1083, which could be used in a similar fashion to the L4/L5 
scenario described previously. 
Halo CMEs observed in \ion{He}{1} 1083 suffer 
from a similar observational 
degeneracy of the large-scale Stokes $U$/$I$ signal, in terms of
the chirality and axial field direction (see Figure~\ref{fig:Halo_CME}), 
at least down to the 1\% polarization level, similarly to the 
previous cases. However, with a polarimetric sensitivity better than 0.5\%, we clearly notice a structural difference between the large-scale Stokes U pattern for  the 
different magnetic topology cases.
Hence, instrumentation with polarimetric precision at the 0.1\% level may be required to unambiguously discriminate both chirality and axial field orientation of a halo CME, or
additional
readily available information about the prominence chirality may be needed to constrain 
the full magnetic field direction.

Based on previous observations and models of the brightness of prominences in
\ion{He}{1} 1083 in the corona by \citet{2015ApJ...802....3M}, we show 
that a space-based coronograph with a 5-cm aperture would be 
able to produce global \ion{He}{1} 1083 polarimetric maps with sufficient
SNR for distinguishing between different magnetic field configurations in the solar corona, as shown 
in Figure~\ref{fig:L45_noise}. Even though previous estimates of the 
\ion{He}{1} 1083 prominence intensity may be
overestimated by an order of magnitude, or biased by significant amounts of instrumental stray 
light, our analysis shows that a robust discernment between different models is still readily 
possible, perhaps with some cost for the temporal and/or spatial resolution of 
the observations.

In conclusion, we have found that the polarization 
signatures of the \ion{He}{1} 1083 line in the solar corona are a promising
diagnostic of the solar coronal magnetic field, with important applications in 
space weather predictive modeling  
\citep[furthering previous work by][]{2016FrASS...3...13D, 2016FrASS...3...20R}.
Future work will extend this modeling effort to compute 
the polarization signals originating from more realistic 3D numerical models of
the solar corona, including a model of solar eruption with self-
consistent prominence condensation \citep{2018ApJ...862...54F} 
and global Alfv\'en-wave heated coronal models 
\citep{2005JGRA..11012226T}.
Based on these realistic models we will be able to introduce a 
significantly higher degree of realism in our computations and to identify more robust examples of the polarization signatures of He I 1083 that can help improve space-weather predictability.

Furthermore, this work is directly translatable to other permitted, Hanle
sensitive, spectral lines in the ultraviolet part of the spectrum 
\citep[as suggested before by][]{2011A&A...529A..12K,2011A&A...532A..70K,2022SoPh..297...96K}, 
supporting the exploitation of the Hanle effect as a powerful and broadly applicable diagnostic of coronal magnetism.

\begin{acknowledgments}
This manuscript greatly benefited from elucidating discussions with Dr.\ Gabriel
Dima (NOAA/CIRES), Dr.\ Nour-Eddine Raouafi (JHU/APL), and Dr.\ Yuhong Fan (NSF NCAR/HAO). The authors would like to thank the referee for the careful reading of the manuscript and the valuable feedback, which  greatly improved the quality of this work. The National Center for Atmospheric Research is a
major facility sponsored by the NSF under Cooperative
Agreement No.\ 1852977. MEM was supported through a UCAR/ASP Postdoctoral fellowship. 
We would like to acknowledge high-performance computing support from the
Cheyenne and Derecho computing clusters \citep{Cheyenne_HPC_System, Derecho_HPC_System} provided by NCAR's Computational and 
Information Systems Laboratory, sponsored by the NSF. CHIANTI is a collaborative project involving George Mason University, the University of Michigan (USA), University of Cambridge (UK) and NASA Goddard Space Flight Center (USA).
\end{acknowledgments}

%

\vspace{5mm}


\software{
\textit{matplotlib}~\citep{Hunter:2007}, 
          \textit{numpy}~\citep{harris2020array}, 
          \textit{scipy}~\citep{2020SciPy-NMeth},
          \textit{mayavi}~\citep{ramachandran2011mayavi}
          }



\bibliography{HeI_10830_coronal_magnetometry}{}
\bibliographystyle{aasjournal}



\end{document}